\documentclass[fleqn,usenatbib]{mnras}

\usepackage[T1]{fontenc}
\usepackage{ae,aecompl}

\usepackage{graphicx}
\usepackage{dcolumn}
\usepackage{bm}
\usepackage{hyperref}
\usepackage{graphicx}
\usepackage{fancyref}
\usepackage{amsmath}
\usepackage{subfigure}
\usepackage{tabularx}
\usepackage{threeparttable}
\usepackage{braket}
\usepackage{mathtools}
\usepackage{stmaryrd}
\usepackage{amssymb}
\usepackage{mathrsfs}
\usepackage{float}
\usepackage{natbib}

\title[Axion Structure Formation I]{Axion Structure Formation I: The Co-motion Picture}

\author[E. W. Lentz et al.]{
Erik W. Lentz,$^{1,3}$\thanks{E-mail: erik.lentz@uni-goettingen.de}
Thomas R. Quinn,$^{2}$
Leslie J Rosenberg$^{3}$
\\
$^{1}$Institut f\"ur Astrophysik, Georg-August Universitat G\"ottingen, 
                 G\"ottingen, Deutschland 37707;\\
$^{2}$Department of Astronomy, University of Washington, 
                 Seattle, WA, USA 98195-1580;\\
$^{3}$Department of Physics, University of Washington,
                 Seattle, WA, USA 98195-1580;
}

\date{Accepted XXX. Received YYY; in original form ZZZ}

\pubyear{2018}

\begin{document}
\label{firstpage}
\pagerange{\pageref{firstpage}--\pageref{lastpage}}
\maketitle

\date{\today}

\begin{abstract}

Axions as dark matter is an increasingly important subject in
astrophysics and cosmology. Experimental and observational searches
are mounting across the mass spectrum of axion-like particles, many of
which require detailed knowledge of axion structure over a wide range
of scales. Current understanding of axion structures is far from
complete, however, due largely to controversy in modeling the
candidate's highly-degenerate state. The series Axion Structure
Formation seeks to develop a consistent model of QCD
axion dark matter dynamics that follows their highly-degenerate nature to the
present using novel modeling techniques and sophisticated simulations. This inaugural paper presents the problem of describing many non-relativistic axions with minimal degrees of freedom and constructs a theory of axion infall for the limit of complete condensation. The derived model is shown to contain axion-specific dynamics not unlike the exchange-correlation influences experienced by identical fermions. Perturbative calculations are performed to explore the potential for imprints in early universe structures.

\end{abstract}



\section{Introduction}
\label{Introduction}

The QCD (quantum chromodynamic) axion is a well-motivated candidate to solve problems in both particle physics and cosmology. The theory of axions originated in 1977 as a result of a scalar-induced axial symmetry over the QCD sector, used to solve the strong (QCD) CP problem of
particle physics \citep{Peccei1977}. The axion particle was proposed
the following year from a spontaneous breaking of that axial
symmetry, solving the strong CP problem dynamically for energies about and below $\lambda_{QCD} \gtrsim 200$ MeV \citep{Weinberg1978,
  Wilczek1978}. Experiments and astrophysical observations have
replaced this Peccei-Quinn-Weinberg-Wilczek axion
\citep{Mimasu2015,Patrignani2016} with more complex and unified
models, pushing the breaking scale ever higher and the axion mass ever
lower. These new theories also have increasingly addressed cosmology \citep{Marsh2016}, where the dark matter (DM) problem has been gaining in prominence.

\begin{figure}[]
\begin{center}
\includegraphics[width=9cm]{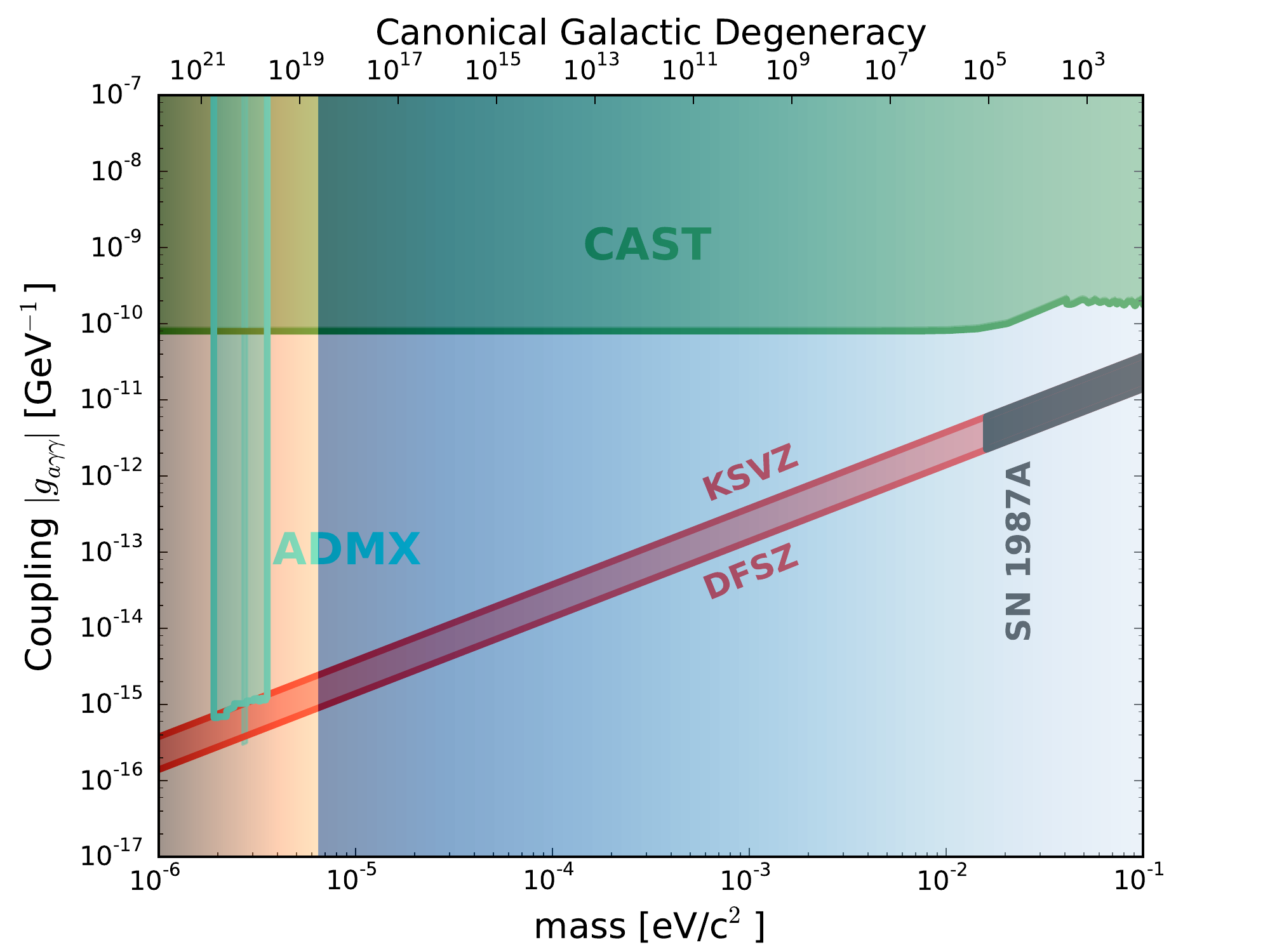}
\caption{The parameter space of axion-like particles over the
  plausible QCD axion DM region, with benchmark theories (red),
  cosmological constraints (orange-to-brown
  gradient, in order of severity) and degeneracy (blue
  gradient). Experimental and observational constraints are also shown
  from ADMX \citep{ADMX2010,ADMX2017}, CAST \citep{Arik2014}, and an
  exemplar of astrophysical bounds SN1987a \citep{Raffelt2008} (cyan,
  green, and gray, respectively). The canonical degeneracy axis is
  calculated from the number of axions contained by a virialized de
  Broglie volume, assuming parameters of $10^8$ M$_{\odot}/$ kpc$^3$
  for the local galactic DM density and $300$ km/s for the velocity dispersion.
}
\label{axionspc}
\end{center}
\end{figure}

As a DM candidate, the QCD axion is highly attractive due to its
well-bounded parameter space of mass and couplings to the standard
model (SM) of particle physics, Fig. \ref{axionspc}. Starting at
milli-eV masses, there is a bound above which the axion would have
been seen in various astrophysical processes
\citep{Raffelt2008,Isern2010,Corsico2012,Viaux2013,Patrignani2016,Marsh2016}
such as anomalous energy transport in SN1987a.  Approaching micro-eV masses, there is a bound below which the maximal (misalignment) creation mechanism would produce more axions than there is DM \citep{Abbott1983,Dine1983,Preskill1983,Patrignani2016}. This lower bound is somewhat soft as axion creation mechanisms can be suppressed in the details of some axion theories \citep{Marsh2016}. The two diagonal lines of Fig.~\ref{axionspc} represent benchmark axion models. KSVZ (Kim-Shifman-Vainshtein-Zakharov) represents a theory where the axion couples to hadrons only \citep{Kim1979,Shifman1980}, and DFSZ (Dine-Fischler-Srednicki-Zhitnitsky) couples to both hadrons and leptons \citep{Dine1981,Zhitnitsky1980} as consistent with grand unified theories. The search window is then given by the region between KSVZ and DFSZ and the lower and upper mass bounds. 

All direct DM searches have in common the need to understand the candidate abundance in the vicinity of their experimental apparatus. Astronomical observations are unable to provide accurate estimates of abundances due to difficulties in extracting local DM contributions from the ephemeral memory of gaseous and stellar motions, though some progress has been made in recent years \citep{Valluri2016,Binney2017}. Instead, simulations that capture the relevant physics of formation are performed to create Milky Way (MW) analogues, and are used to estimate DM populations. Methods for performing these simulations are already quite sophisticated, capable of detailing the formation of cosmological structure down to the inner workings of galaxies. Often starting from cosmic microwave background (CMB)-motivated initial conditions, simulations of structure formation must accurately describe the violent process of collapse, collision, and equilibration of matter into stars, galaxies, halos, and other forms.

There are several critical points in the early universe that significantly impact the modern relic axion state. The first centers on the relative placement of the  spontaneous symmetry-breaking scale of the axion's parent scalar field, $f_a$, and the inflation scale $f_I = T_I/2 \pi$ where $T_I$ is the associated cosmological radiation temperature. In brief, if the PQ symmetry is unbroken during inflation, meaning that $f_a< T_I/ 2 \pi$, then the axion's parent scalar field has zero vacuum expectation value (VEV), leaving the angular axion field randomly distributed over each Hubble patch when the $f_a$ is reached after inflation. Such a distribution results in the large average misalignment angle $\left< \phi^2_{a,i} \right> = \pi^2/3$. If the PQ symmetry is broken during inflation, $\phi_{PQ}$ varies on scales above the Hubble volume post-inflation. This effectively makes $\phi_{PQ}$ uniform in the visible universe and close to a free parameter of the model.

Issues occur with each inflation possibility. The randomness of
$\phi_{PQ}$ in the pre-PQ breaking inflation case gives rise to topological defects and seeds other structures that may prove difficult for local searches such as haloscope experiments, as much of the mass may exist in compact objects such as mini-clusters \citep{Kolb1993, Tinyakov2016, Fairbairn2017}. Post-PQ breaking inflation, which produces relatively lighter axions, introduces a degree of anthropic reasoning to the state of the visible universe. A narrow range of misalignment angles are allowed to reproduce the observed DM density, severely constraining the parameter. Even with inflation-enforced flatness, striking the necessary balance of axion DM with other components makes even a semi-random axion phase precarious. 

The second critical point occurs when the post-symmetry-breaking
post-inflation misalignment angles leave the freeze-out state. This occurs for the QCD axion during the QCD phase transition. The condensation of the quark-gluon plasma creates an effective potential for the axions such that the potential minimum is located at a point of vanishing CP violation. There
are several popular models for producing cosmological axions,
including the topological defect decays from phase windings, decays from
parent particles, thermal populations, and vacuum realignment. We
restrict ourselves to the defect-free realignment mechanism at this time, though
the other possibilities may be explored in future work.

The misalignment process is very clean, creating nearly homogeneous
axions within a Hubble patch, and restricting the scale of spatial
oscillation to be of order the horizon size or larger in the absence
of topological defects. The initial axion state is incredibly
degenerate, with the canonical critical temperature estimated to be of
$O(100GeV)$ for micro-eV axions, far above the QCD axion mass \citep{Banik2015c}. It is reasonable to assume that the axions inhabit some form of Bose-Einstein condensate (BEC) from the expected degeneracy alone.

Axion DM dynamics go through several stages after creation. Radiation
gravity and the potential of the axion field dominate axion interactions early after creation,
respectively equilibrating with the cosmological radiation
perturbations or inducing violent collapse in areas of large
over-density \citep{Redondo2013,Gorghetto2018,Vaquero2018}.  Axion self-interaction reduces quickly with further cosmological expansion, leaving  dynamics dominated by axion-conserving interactions from environmental and self gravity. QCD axion DM interactions are dominated by gravity by the time of the matter-radiation transition; this letter concentrates on structure formation after that transition.

The state of axion structure formation after self-interactions consists primarily of two viewpoints. The first finds sufficiency in a mean-field description of a condensed axion fluid while the second holds that the full quantum description is necessary to describe the axion field. Relic axions created under the misalignment mechanism easily satisfy the canonical requirements of a thermal BEC. It therefore seems reasonable to assume that every axion is in essentially the same state. Proponents of the mean field description motivate simplification along this single-state constraint over every axion. Implementing such a mean field theory (MFT) on relic axions is straightforward and attractive. In the limit of ultra-light axions (ULAs), $m_a \sim (few) \times 10^{-22}$ eV, the de Broglie wavelength scale enters the galactic scale \citep{Arvanitaki2010,Schive2014,Veltmaat2018}. QCD axions are at least sixteen orders of magnitude heavier and will not see outright wave diffusion effects on galactic scales. This reduces the dynamics to a classical pressure-less fluid, the same used in the cosmological standard model $\Lambda$CDM. The overlap of axion and standard cold DM (CDM) models would relegate the detection of axions to direct or otherwise non-gravitational methods.

The second portion of the axion structure formation community proposes that quantum effects extend far beyond the de Broglie scale. This invites the possibility that QCD axion DM may produce observable signatures unique to axions  on galactic scales. Models built on requirements of coherent energy and angular momentum conservation at the quantum level produce halos where axions collapse as a series of cold phase-space sheets, drawn preferentially towards the galactic plane by their rotational properties. The resulting post-collapse halo is found to contain of a series of caustic rings, the effect of $D_{-4}$ catastrophes \citep{Sikivie1999a,Sikivie1999b,Duffy2008,Chakrabarty2017}.
It is also claimed that our own solar system is close to one of these caustics of a fitted MW axion halo. Cold flows feeding into the outermost rings would contribute to the local axion energy spectra, significantly influencing the results of terrestrial search experiments. Proponents of caustic rings  appreciate the compactness of the MFT description, but believe that it does not adequately maintain the quantum nature of the degenerate Bose system. The fault with the MFT simplification is reported to lie in the mis-estimation of the relevant timescales of gravitational equilibration, leading to an imbalance in state occupation \citep{Banik2015b,Erken2012}. The more complete axion models used in the gravitational thermalization research are implemented at the level of individual axion quanta and unfortunately cannot be extended to cosmological scales.

Non-scalability notwithstanding, a more complete quantum description
of axions than MFT is required for several reasons. First, current axion MFTs do not track inter-axion correlations. Standard
condensate MFTs use a single-particle Fock decomposition of the axion
solutions combined with an assumption of separability between axion
states to produce a compact single Slater permanent representation of
the system. However, gravity is an infinite range, un-screenable, and
highly-correlated interaction that makes such a separation in the gravitational solution space unfeasible. It is likely that the constraint of axion infall models to such confined states introduce unphysical forces to the axions' dynamics. 

Further, the expected energy level of a cosmological system of axions
is nowhere close to its ground state, nor is it expected to reach
thermal equilibrium. Classically, self-gravitating systems collapse to
a virialized quasi-equilibrium, which is known to be distinct from a
thermalized fluid by phenomena such as the gravi-thermal catastrophe
\citep{BT2008}. Gravitationally-virialized fluids are also expected to
maintain knowledge of their infall history, which may be valuable to
astronomers and dark matter researchers \citep{Valluri2016}. Structures
formed with non-local Newtonian gravity may conceptually be quite
different than the canonical thermodynamic approach, requiring
rigorous definitions of condensation and condensates. The ignoring of correlations in current axion MFTs also extends to the interactions, smoothing out potential two-body forces. There is therefore a need for a beyond-MFT model of axion dark matter, capable of tracking axions in and out of equilibrium.

It is the objective of this series to provide a model of axion
structure formation sufficient to resolve axion-specific physics on
astronomical scales and determine the extent of unique
structures. This first letter derives a new model of QCD axion DM
structure formation in the condensed limit and calculates its
potential imprints on the early visible universe. The remainder of
this paper is structured as follows: Section~\ref{ALU} describes the
context for axion structure formation after the QCD phase
transition and into modern times; Section~\ref{New} introduces the new
continuum model of structure formation for a system of perfectly
condensed self-gravitating bosons;  Section~\ref{AS}  performs calculations of condensed axion structure formation in the small-perturbation limit, exploring the extent of unique axion structures in the early universe; Section~\ref{Discussion} discusses the implications of the new model and the perturbative results; and Section~\ref{Summary} summarizes the results and outlines the scope of upcoming letters.

\section{Axion Physics after the QCD Phase Transition}
\label{ALU}

Post-QCD phase transition axions are well-described by a single
quantum scalar field operator ($\hat{\phi}$) defined over a dynamic
3+1 pseudo-Riemannian manifold ($M^{3+1}$) governed by classical
Einstein's equations. Such a full description is extremely accurate
but excessive for common cosmological structures, so several
simplifications are made in the interests of computability. On the
largest scales, the geometry is taken to be nearly homogeneous,
isotropic,
while the smaller scales of interest are taken to satisfy weak-field
conditions \citep{MTW}. The space is then found to evolve under a
Friedmann-Lemaitre-Robertson-Walker (FLRW) cosmology with small
geometric perturbations. Cosmological measurements favor the flat
variety of FLRW cosmologies \citep{Planck2018Cosmo}, requiring a critical energy density among all gravitating species.  Einstein's equations then decouple in orders of metric perturbations, with the background flat FLRW cosmology evolving under the horizon-scale-averaged stress-energy
\begin{align}
&H^2 - \frac{\Lambda c^2}{3} = \frac{8 \pi G}{3} \bar{\rho}, \\
&2 \frac{ \ddot{a}}{a} + \left(\frac{\dot{a}}{a}\right)^2 - \Lambda c^2 = -\frac{8\pi G}{c^2}\bar{p},
\end{align}
where $\bar{\rho}$ is the total expected mean co-moving energy density, $\bar{p}$ is the total expected mean pressure, $a(t)$ is the FLRW scale factor, $H = \dot{a}/a$ is the Hubble flow, and the dots represent derivatives in physical time as measured by a co-moving observer. The next order terms obey a parabolic potential-like equation.
\begin{equation}
\Delta \Phi = \frac{4 \pi G}{a c^4} \rho^{(1)}
\end{equation}
where $\Delta = \vec{\nabla}^2$ is the flat spatial Laplacian,
$\rho^{(1)}$ is the deviation in the frame's energy density from the
mean, and the Newtonian potential is related to the co-moving geometry
via $\Phi = - h_{00}/4$ where $h_{00}$ is the diagonal time component
of the co-moving metric perturbation from FLRW. Relativistic gravity
as a classical theory can be interpreted as averaging over quantum length scales. Free fundamental particles are associated with two quantum length scales,
Compton and de Broglie, though only Compton is covariantly meaningful;
the de Broglie lengths parameterize frame-dependent corrections from
wave kinematics. As the fundamental length scale of a particle, the Compton length
determines the limit by which a light-like vector field may distinguish space-like structure. The Compton scale is on the order of centimeters for QCD axion DM. This is the smoothing length that shall be used to evaluate the axion gravitational potential.

The effective axion equations of motion are also derived from a fully covariant description. Describing the axion system as a second-quantized operator, the axion field's governing equations are characterized by a Lagrangian operator of the form
\begin{equation}
\hat{\mathcal{L}} = \left(\nabla_{\mu} \hat{\phi}^{\dagger} \right) g^{\mu \nu} \left(\nabla_{\nu} \hat{\phi} \right) - V(\hat{\phi}) 
\end{equation}
where $\nabla_{\mu}$ is a covariant derivative over $M^{3+1}$ and $V$ is the axion interaction potential. The axion field operator is of Bose type, meaning that it commutes under particle exchange. The field obeys the Klein-Gordon equation over effective fields in the semi-classical limit
\begin{equation}
\nabla^{\mu} \nabla_{\mu} \hat{\phi} + \left(\frac{m_a}{\hbar}\right)^2 \hat{\phi} - \frac{\lambda}{3!} |\hat{\phi}|^2 \hat{\phi} = 0
\end{equation}
where the fourth order expansion of the axion potential is used instead of the full cosine axion potential 
\begin{equation}
V(\hat{\phi}) = \frac{m_a^2}{\hbar^2} |\hat{\phi}|^2 - \frac{\lambda}{4!} |\hat{\phi}|^4 + O(|\hat{\phi}|^6)
\end{equation}
and $\lambda = \propto f_a^{-2}$ is the coupling strength of the four-field coupling strength. Evaluating the covariant derivatives in the linear approximation, the derivative term of the Klein-Gordon operator becomes
\begin{align}
g^{\mu \nu} \nabla_{\mu}  \nabla_{\nu} \hat{\phi} &= a^{-2} \square \hat{\phi} + 3 a^{-2} H \partial^t \hat{\phi} \nonumber \\
&- a^{-2} \left(h^{-1}\right)^{\mu \nu} \partial_{\mu} \partial_{\nu} \hat{\phi} - 3 a^{-2} H \left(h^{-1}\right)^{0 \nu} \partial_{\nu} \hat{\phi} \nonumber \\
&- a^{-2} h \partial^{0} \hat{\phi} + \frac{1}{2} a^{-2} \partial^{\nu}(h) \partial_{\nu} \hat{\phi} +O(h^2)
\end{align} 
where $\square = - \partial_t^2 + \Delta$ is the flat d'Alembert operator, and all index contraction, raising and lowering, are performed using the Minkowski metric ($\eta$) in agreement with the previously-mentioned weak field formalism \citep{MTW}.

The effective Newtonian limit eliminates the lowest order terms and is consistent with non-relativistic fields in the co-moving frame, allowing for the non-relativistic field expansion 
\begin{equation}
\hat{\phi} = \frac{\hbar }{\sqrt{2} m} \left( \hat{\psi} e^{-i m_a t/\hbar} + \hat{\psi}^* e^{i m_a t/\hbar } \right)
\end{equation}
Substituting this form of the field operator, and taking the Newtonian limit for gravity, $h_{00} = - 4 \Phi$, the real time Klein-Gordon operator equation reduces to
\begin{align}
& \hat{\psi}^*\left(\partial_t - \frac{3}{2} H  \right)\hat{\psi} +\hat{\psi}^* \frac{\hbar^2 \vec{\nabla}^2}{2 a^2 m_a} \hat{\psi} + m_a \Phi |\hat{\psi}|^2 \nonumber \\
&- \frac{\lambda \hbar^4}{3! m_a^4} |\hat{\psi}|^4+ \{h.c.\} + (\text{quickly oscillating terms}) \nonumber \\
&= O(h^2, m^2/\hbar^2,m h/\hbar) 
\end{align}
The quickly oscillating terms are ignored as their actions evaluate to
$0$ over first-order timescales. The terms of the effective axion
action can be easily related to the terms of the field Hamiltonian
operator as the action measure includes a factor of $\sqrt{-g}= a^2 +
O(h)$:
\begin{align}
\hat{H}_f &= \hat{K} + \hat{U} + \hat{V} \\
\hat{K} &=  - \hat{\psi}^* \frac{\hbar^2 \vec{\nabla}^2}{2 a^2 m_a} \hat{\psi} \\
\hat{U} &= \frac{\lambda}{4!}|\hat{\psi}|^4 \\
\hat{V} &=  \hat{\psi}^*  m_a \Phi \hat{\psi}
\end{align} 
which confirms that $\Phi$ is indeed the canonical Newtonian gravitational potential. 

The many-body axion Schr\"odinger operator equation can now be written in the comoving parameterization
\begin{equation}
i \hbar \partial_t \hat{\psi} = - \frac{\hbar^2 \vec{\nabla}^2}{2 a^2 m_a} \hat{\psi} + \frac{\lambda}{4!}|\hat{\psi}|^2 \hat{\psi} +  m_a \Phi \hat{\psi}
\end{equation}
This is the operator equation governing the quantum axion field in the presence of classical gravity, which serves as a robust starting point for much of axion structure-formation models after the QCD phase transition.

Interactions during the remainder of the radiation era are dominated by gravitational perturbations from relativistic species and the axion point interaction. 
Structure formation of axions are linear so long as axion perturbations remain small during appreciable self-interaction. After a time the axions diffuse beyond appreciable self-interaction, the mean free path exceeds the Hubble radius, and their number becomes a conserved quantity.

The expansion rate and diffuse gravitational potentials keep the axions in the perturbative regime during the remainder of radiation domination. The field operator picture is still sensible during this time of point-wise and separable potentials. Even so, there is some sub-dominant but increasingly important physics at work. Inter-axion gravitation connects every axion to every other axion via the long-range un-screenable Coulomb potential
\begin{equation}
V_{grav}(\vec{x}_i,\vec{x}_j) \cong -\frac{4 \pi G m_a^2}{|\vec{x}_i-\vec{x}_j|}
\end{equation}
for well-separated axions in this limit of gravity. Such a highly-correlated interaction is thought to maintain the axion condensate through ``gravitational thermalization'' \citep{Erken2012}, though this may be somewhat irrelevant as normal dispersion processes are not anticipated to push the DM out of the canonical condensate phase.  

As matter energy density establishes equality and then dominance over radiation, axion self-gravity becomes the primary source of structure-formation (SF) dynamics. Maintaining the field operator form of the system becomes more difficult during this process as the Fock representation of the many-axion system becomes more tenuous. Degradation in the separability and perturb-ability of the Hamiltonian at best requires the use of a longer chain of Fock states.
Regardless, one can expect computations over a separable set of axions to become untenable as we enter the DM and dark energy (DE) dominated eras. MFTs are then seen to fail in describing axion SF in the late universe as they are constrained from tracing the correlated states.
The next section describes a new model of axion SF that  more fully explores the state space.

\section{A New Model of DM Axions}
\label{New}

This section follows similar path to \citet{Lentz2018a}. Embellishment
on the finer points of Schr\"odinger and Wigner methods in the context
of axion self-gravitation are given in Appxs.~\ref{MBS}, \ref{Wigner},
\ref{DFInt}.

\subsection{Many-Body Quantum Axions as a Many-Body Schr\"odinger Equation}
\label{ManyBody}

While in principle correct, the second quantization formalism used by
many axion MFTs is not well-suited for axion structure formation. It
encourages a separable decomposition among axions not supported by
their dynamics, and supplies a superfluous freedom to
create/annihilate said separated particles. The technique presented
here instead restricts its scope by falling back to first
quantization, made possible by conserved axion number. Rephrasing the
field theory into a quantum mechanical approach, the Schr\"odinger picture of Hilbert space uncovers a many-body wave-tracing evolution equation. Evaluating the dynamical equation over an arbitrary element of the many-body solution space gives the form
\begin{equation}
 i \hbar \partial_t \Psi(\vec{x}_1,...,\vec{x}_m;t) = \hat{H} \Psi(\vec{x}_1,...,\vec{x}_m;t)
\end{equation}
where $\Psi$ is the many-body axion wave-function and $\hat{H}$ is the many-body Hamiltonian operator on wave-function space.  Note that the symbol for the first-quantized many-body Hamiltonian operator $\hat{H}$ is similar but distinct from the Hubble flow $H$. The wave-function also inherits an exact exchange symmetry due to the commuting nature of the axion field operator. 
\begin{align}
\Psi \left( \vec{x}_1,...,\vec{x}_i,...,\vec{x}_j,...,\vec{x}_N;t \right) &= + \Psi \left( \vec{x}_1,...,\vec{x}_j,...,\vec{x}_i,...,\vec{x}_N;t \right) \nonumber \\ 
& \forall i,j
\end{align}
The many-body Hamiltonian $\hat{H}$ may be broken down into recognizable kinetic and potential energy parts
\begin{equation}
\hat{H} = \hat{T} + \hat{V} = \sum_i^M \frac{- \hbar^2 \nabla_i^2}{2 a^2 m_a} + \sum_i \Phi_i 
\end{equation}
where $\Phi_i$ is the gravitational potential on the i-th axion.

The gravitational interaction requires further clarification. Strictly speaking the gravitational potential may come from multiple massive and radiative species. Considering only massive species like axion DM, stars, atomic gas, etc., the potential may be broken up linearly, using the Coulomb gauge choice where the potential vanishes when infinitely far from all sources:
\begin{align}
\rho &= \rho_{a} + \rho' \text{ and}\\
\nabla^2 \Phi &= 4 \pi G \rho = 4 \pi G \left( \rho_{a} + \rho' \right) 
\end{align}
implies
\begin{align}
\Phi &= \Phi_{a} + \Phi',
\end{align}
where
\begin{align}
\nabla^2 \Phi_a &= 4 \pi G \rho_a \\
\nabla^2 \Phi' &= 4 \pi G \rho' 
\end{align}
and all potential terms not sourced by axionic densities are consolidated into $\rho'$ and $\Phi'$ and referred to hereafter as ``baryons''. These objects are highly classical and require no further gravitational consideration than has already been made in the literature. The inter-axion gravitational Poisson equation may be inverted using standard Greens function techniques
\begin{equation}
\Phi_a \left( \vec{x} \right) = - \int d^3 x' \frac{4 \pi G\rho_a(\vec{x}')}{|\vec{x}-\vec{x}'|}
\end{equation}
From the previous discussion of stress-energy smoothing it is determined that the axion density should be sampled on the Compton scale. The potential of a collection of Compton-limited axions is then
\begin{equation}
\Phi_a \left( \vec{x} \right) = -\sum_i^N \int d^3 x' \frac{4 \pi G \rho_C(\vec{x}_i)}{|\vec{x}-\vec{x}_i|}
\end{equation}
where the sum is over every axion and $\rho_C$ is the axion's
Compton-limited distribution profile. The result looks similar to
electrons interacting electro-statically in simple atomic and
molecular systems. The gravitational potential for an axion from other axions is then
\begin{equation}
\Phi_{a,j} = -\sum_{\substack{i ,\\ i \ne j}}^N \int d^3 x' \frac{4 \pi G \rho_C(\vec{x}_i)}{|\vec{x}_j-\vec{x}_i|}
\end{equation}
where the contribution of the axion onto itself is expressly omitted. For the remainder of the paper we shall use the shorthand definition of the inter-axion gravitational potential
\begin{equation}
\phi_{ij} = \phi(\vec{x}_i,\vec{x}_j) = -\int d^3 x' \frac{4 \pi G \rho_C(\vec{x}_i)}{|\vec{x}_j-\vec{x}_i|}
\end{equation}
The form of the many-body Hamiltonian may now be rewritten
\begin{equation}
\hat{H} = \sum_i^N \frac{- \hbar^2 \nabla^2}{2 a^2 m_a} + \frac{1}{2} \sum_{i \ne j}^N \phi_{ij} + \sum_i^N \Phi'(\vec{x_i}) \label{MBHam}
\end{equation}
where the inter-axion potential sum is over both $i$ and $j$ indexes such that $i\ne j$.

\subsection{Solutions of the Many-Body Equation}
\label{MBsolns}

A more detailed version of this section may be found in Appx.~\ref{MBS}, which outlines the context of the many-body bosonic Schr\"odinger equation, determines its implicit and explicit symmetries, and derives solutions. Presented here is a class of useful solutions to motivate later derivations.

The axion DM Hamiltonian is far from separable across single particle
lines, making a Fock space representation and reduction of states
untenable. In the case of no external potentials, there exists a more concise form to solutions of the interchange-symmetric Schr\"odinger equation, which centers on inter-particle correlators. The center-of-mass solutions of the system are then spanned by fully symmetric combinations of inter-particle correlators
\begin{equation}
\Psi \left( x,t \right) \in sp \left( \left\{ perm \left( \prod_{ij } \psi_{\alpha_{ij}} \left(\vec{x}_i - \vec{x}_j,t \right) \right) \right\}_{\alpha \in A} \right)\label{cohstates}
\end{equation}
where $sp()$ gives the linear span of a collection of functions, $perm()$ is the permanent operation, $A$ is the set of many-body solution indexes, $\hat{\psi}_{a}$ is a solution to the single-body equation
\begin{equation}
i \hbar \partial_t \psi_{a} = - \frac{\hbar^2}{2 a^2 \mu} \nabla^2_{\left( \vec{x}_i - \vec{x}_j \right)} \psi_{\alpha} + \phi\left( \vec{x}_i - \vec{x}_j \right) \psi_{a}
\end{equation}
  and $\mu$ is a particle reduced mass and scales as $1/N$, thereby giving a naturally-extended potential length scale to the inter-particle correlations. The inter-particle potential's length scale, the number of particles $N$, and the global reach of permutation (anti-)symmetry are therefore key to the development of the system's structure. This correlator form explores the solution space more efficiently than the single-particle expansion natural to Fock representations, especially in the presence of highly-correlated interactions. The utility of this form lies in the implicit embedding of exact exchange symmetry and the effects of that constraint on possible actions of the system; this is crucial in the expression of physics beyond the mean field. Only condensed configurations are considered for the remainder of this paper, where condensed here is taken to mean occupation of a single basis element of Exp. \ref{cohstates} with $\alpha_{ij} = \alpha$. More detailed considerations of condensation, including a generalized density-of-states derivation conforming to early axion dynamics, will appear in a later paper of this series.

\subsection{Distillation to a Condensed Distribution Function}
\label{}

Condensing solutions of the many-body axion Schr\"odinger equation to
the salient degrees of freedom is next. Applying the Runge-Gross
theorem \cite{Runge1984,TDDFT1984}, which proves the existence of an
injective mapping from the potentials of a many-body quantum
mechanical system to a single-body density when given an initial many-body
wave-function, reveals that the only important degree of freedom for
the identical system is its density, $\rho$. The Runge-Gross theorem
therefore reveals the possibility of reducing dimensionality without
loss of generality of the axion state. Reduction of a cosmological volume of axions to a single body's dynamics is therefore possible. Our treatment aims for such a description.

 The approach pursued here phrases the density in terms of distribution functions (DFs), which track the density of system attributes through phase space. The many-body wave-function can be reduced into a phase-space pseudo-DF via a many-body Wigner transform. At the classical level ($\hbar \to 0)$, the transform produces a DF governed by a collision-less Boltzmann equation
\begin{align}
&\partial_t f^{(N)} + \sum_i^N \frac{\vec{p}_i}{a^2 m_a} \vec{\nabla}_i f^{(N)} - \sum_i^N \vec{\nabla}_i \Phi' \cdot \vec{\nabla}_{p_i} f^{(N)}  \nonumber \\
&- \frac{1}{2}\sum_{i \ne j}^N \vec{\nabla}_i \phi_{ij} \cdot \vec{\nabla}_{p_i} f^{(N)} = O(\hbar) \label{DFEOM}
\end{align}
where $f^{(N)}$ is the N-body DF. See Appx. \ref{Wigner} for more detail on Wigner transforms and their properties. At this point neither the specific dynamics nor exchange conditions on the state are enforced, implying that such a Boltzmann equation may also by applied to analogous classes of fermionic or mixed systems.

Reducing the degrees of freedom to a single body is now a matter of integration. Projecting from the full N bodies to a two-body DF is straightforward, covered in detail in Appx. \ref{DFInt}, but reduction from two to one body is more complex due to the presence of two-body correlations. The normal ``molecular chaos'' approach to distribution theory would lead one to take a near-uncorrelated form of the distribution $f^{(2)} = f^{(1)} \otimes f^{(1)} + g$. The correlation function of this form is often restricted to  $1/N$ scaling by local considerations of classical two-particle scattering \cite{Fokker1914,Kolmogoroff1931}. However, the condensed state solutions  from Sec.~\ref{MBsolns} implies no such scaling due to the global reach of the permutation condition. Instead, the extremal nature of the condensate is used to construct a correlation optimization problem. Defining the two-body correlation function by decomposing the two-body DF as 
\begin{equation}
f^{(2)}\left(w_1,w_2,t\right) = \tilde{g} \times f^{(1)}(w_1,t) f^{(1)}(w_2,t),
\end{equation}
correlation functions of the extremal case of a condensed Bose fluid are found with minimal computation 
\begin{equation}
\tilde{g} = \frac{C-\lambda_1 f_+}{1+ \lambda_2 f_+}
\end{equation}
where $f_+$ is the symmetrized single-body DF, and $\lambda_1$ and $\lambda_2$ are integrals of motion, and $C$ is constrained by the correlation present in the initial configuration; details are available in Appx. \ref{DFInt}. The resulting single-body equation of motion may now be written 
\begin{align}
&0=\partial_t f^{(1)} + \frac{\vec{p}}{a^2 m_a} \cdot \vec{\nabla} f^{(1)} - \vec{\nabla} \Phi' \cdot \vec{\nabla}_p f^{(1)} - \vec{\nabla} \bar{\Phi} \cdot \vec{\nabla}_p f^{(1)} \nonumber \\
& - \frac{N-1}{N} \int d^6w_j \vec{\nabla} \phi_{1j} \cdot \vec{\nabla}_{p} f^{(1)}(w_1,t) f^{(1)}(w_j,t) \nonumber \\
&\times \frac{C - 1 -(\lambda_1+\lambda_2) f_+}{1+\lambda_2 f_+} \label{boseboltz}
\end{align}
where $\bar{\Phi}$ is the axion Newtonian gravitational potential, classically averaged 
\begin{equation}
\bar{\Phi} = N \int d^6 w_2 \phi_{12} f^{(1)}
\end{equation}
The extra-classical physics is a direct result of the exchange symmetry shared by each axion pair and the non-trivial correlations between axions; this influence is referred to as the exchange-correlation (XC) interaction. The effect of the XC pseudo-forces is almost classical, in that the removal of any one axion does little to change the result.

While compact, the form of Eqn. \ref{boseboltz} is not simple to compute for many calculations in structure formation. Conveniently expanded forms of the new Boltzmann-like equation and other equations relevant for perturbative calculation are provided in Appx.~\ref{EXP}.

\section{Early Axion Structures}
\label{AS}

This section investigates the presence of unique structure in early axion structure formation. ``Early'' here is taken to be synonymous with linearly perturbative over a homogeneous and isotropic background, such as in most radiation-dominated to matter-dominated recombination era models of SF.

Perturbative dynamics in the hydrodynamic limit are well recorded in the literature, including references \citet{BT2008} and \citet{MvdBW2010}. The observables of fluid density and velocity are tracked in this demonstration, though higher moments are straightforward to compute. The density takes the classical definition of density perturbations in a co-moving frame
\begin{equation}
\rho(\vec{x}) = M \int d^3 p f (\vec{x},\vec{p}, t) =  \bar{\rho} a(t)^3 \left(1 + \delta(\vec{x},t)\right)
\end{equation}
where $M$ is the enclosed mass and $\bar{\rho}$ is the average density over the Hubble volume. The fluid velocity takes on a similar definition
\begin{equation}
\left< \vec{v}(\vec{x}) \right> = \frac{1}{\rho} \int d^3 p  f (\vec{x},\vec{p}, t)\vec{p}
\end{equation}
The fluid is also considered to be nearly static in the co-moving frame, with flows of order $\delta$.

The linearized hydrodynamics of CDM are derived first. The co-moving CDM Boltzmann equation
\begin{equation}
\partial_t f^{(1)} + \frac{\vec{p}}{m a^2} \cdot \vec{\nabla} f^{(1)} - \vec{\nabla} \Phi' \cdot \vec{\nabla}_p f^{(1)} -\vec{\nabla} \bar{\Phi} \cdot \vec{\nabla}_p f^{(1)} = 0
\end{equation}
is expanded in orders of density and velocity
perturbations. Expectations for the zeroth and first $\vec{v}$ moments of the distribution are found to have dynamics
\begin{align}
0^{th}&: \partial_t \delta + \frac{1}{a} \vec{\nabla} \cdot \left\{\left(1+\delta \right) \left< \vec{v} \right> \right\}  = 0 \label{ncons}\\
1^{st}&: \partial_t \left\{\left(1+\delta \right) \left< \vec{v} \right> \right\} + H \left(1+\delta \right) \left< \vec{v} \right> + \frac{\left(1+\delta \right)}{a} \vec{\nabla} \left(\Phi' + \Phi \right) \nonumber \\
&+ \frac{1}{a} \vec{\nabla} \cdot \left\{ \left(1+\delta \right) \left< \vec{v} \otimes \vec{v} \right>  \right\} = 0 \label{pcons}
\end{align}
giving  density and momentum conservation laws respectively. An
Euler-like form may be derived through a combination of
Eqns.~\ref{ncons} and \ref{pcons}:
\begin{align}
&\partial_t \left< \vec{v} \right> + H \left< \vec{v} \right> + \frac{1}{a} \left(\left< \vec{v} \right> \cdot \vec{\nabla} \right) \left< \vec{v} \right> + \frac{1}{a} \vec{\nabla} \left(\Phi' + \Phi \right) \nonumber \\
&+ \frac{1}{\left(1+\delta \right) a} \vec{\nabla} \cdot \left\{ \left(1+\delta \right) \left( \left< \vec{v} \otimes \vec{v} \right> - \left< \vec{v} \right> \otimes \left< \vec{v} \right> \right) \right\} = 0,
\end{align} 
where the last term disappears as cold dark matter has vanishingly small velocity dispersion. The fluctuation equation of motion is derived by combining the Euler form and matter conservation again, and to first order looks like a damped harmonic instability
\begin{equation}
\partial^2_t \delta + 2 H \partial_t \delta = 4 \pi G \bar{\rho} \left( \delta + \delta' \right) + O(\delta^2)
\end{equation}
where $\delta'$ is the energy density fluctuation from complementary species. This gives the expected collapse on all concerned scales, in the limit of no back reactions.

Calculating perturbative Bose hydrodynamics follows very similarly. The lowest order XC effects modify the perceived depth of the self-gravitational well, which is analogous to modifying the observable cosmological dark matter density, but only within the DM species. The fluctuation equation of motion is derived to be
\begin{align}
&\partial^2_t \delta + 2 H \partial_t \delta = 4 \pi G \bar{\rho} \left( C \delta + \delta' \right) \nonumber \\
&- 4 \pi G \bar{\rho} \frac{\lambda_1 + \lambda_2}{2 W} 3 \delta +
  O(\lambda_2^2,\delta^2), \label{forceamp}
\end{align}
where the $\lambda$s are evaluated at the initial spectrum and $W$
represents the phase space volume over which the $\lambda$s are calculated. The used expansion of the 
correlation function can be found in Appx.~\ref{EXP}. MFT is reached
in the homogeneous limit, meaning to zeroth order $C = C_0 = 1$ and
$\lambda_i = \lambda_{i,0} = 0$. Strictly speaking, the XC has no
impact on first order dynamics as the non-trivial components of $C$ and the $\lambda$s enter at first order in perturbations. However, an inflated XC impact will be
entertained here for the purposes of demonstration. A
simple relation between the initial correlation and the constraints is found at first order in perturbations
\begin{equation}
C-1 = C_1 = \frac{\lambda_{1,1} + \lambda_{2,1}}{W}. \label{pertC}
\end{equation}

\begin{figure}[]
\begin{center}
\includegraphics[width=9cm]{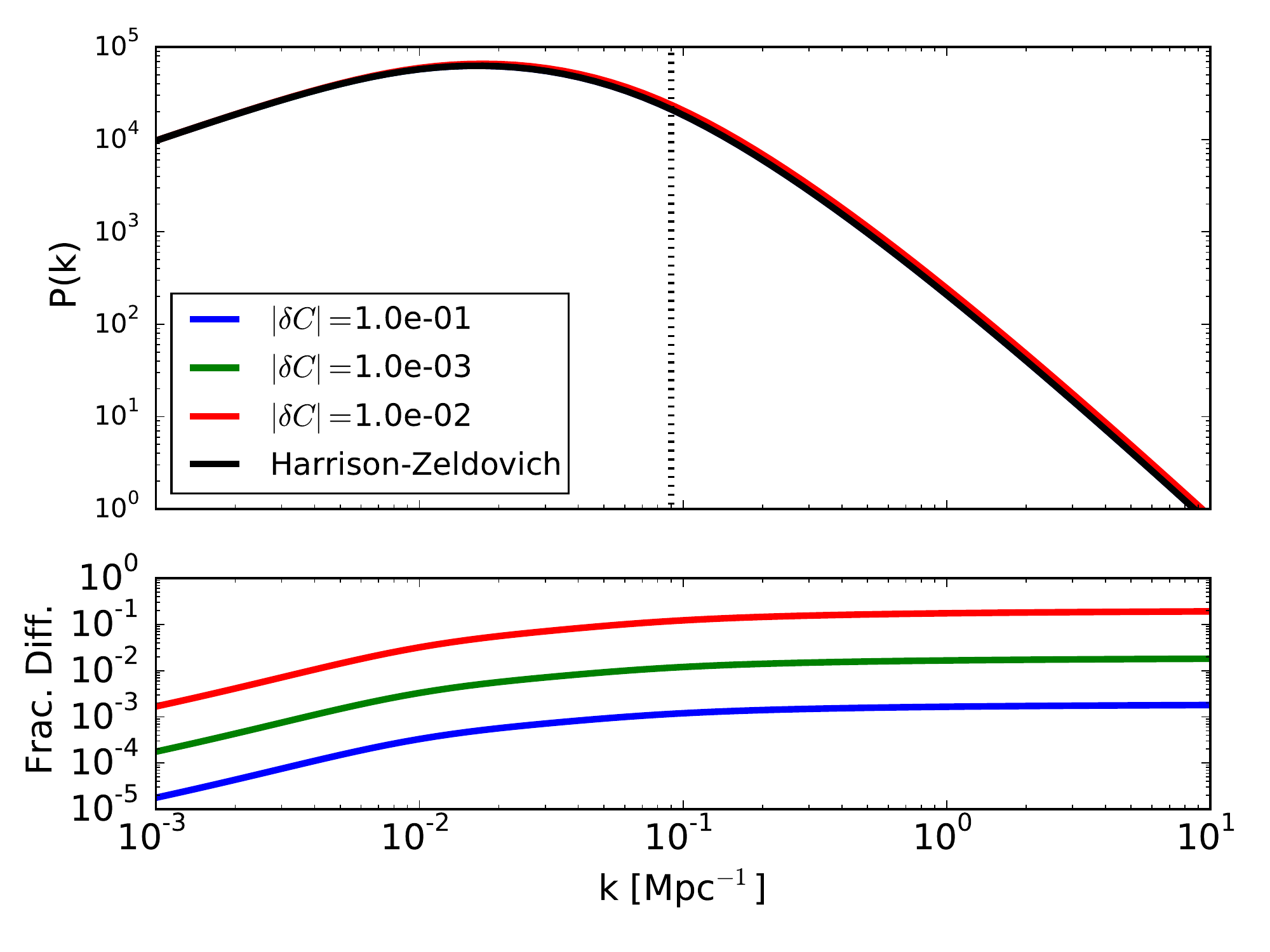}
\caption{Perturbation power spectra evolved to $z=0$. The upper panel shows the perturbation-theory estimated power spectra evolved under both CDM and Bose dynamics. The CDM spectra is consistent with that found in \citet{Spergel2007}. Bose spectra are calculated using the same relic density as CDM, plus the gravity amplification derived in Eqns. \ref{forceamp} and \ref{pertC}. Deviations in the Bose initial correlations $\delta C$ vary over multiple orders in magnitude. Spectra are shown well into the non-linear regime, as demarcated by the vertical dashed line. The lower panel shows fractional  power differences between the CDM and Bose spectra.}
\label{CMBpowspec}
\end{center}
\end{figure}

\begin{figure}[]
\begin{center}
\includegraphics[width=9cm]{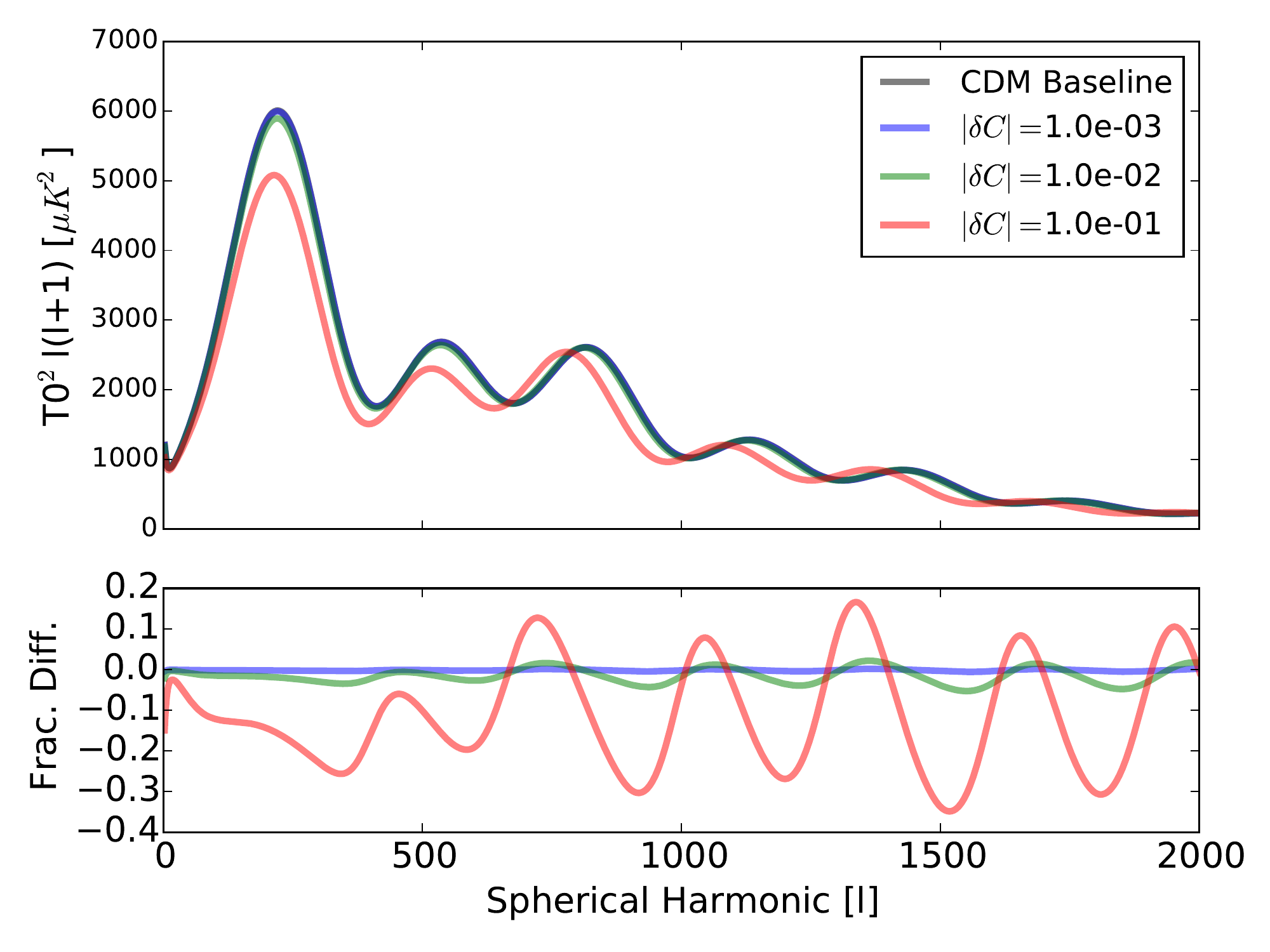}
\caption{Perturbative baryon angular power spectra at recombination as influenced by background DM. The top panel shows the perturbative angular power spectra induced into the baryonic component by both CDM and Bose background distributions. Deviations in the Bose initial correlations $\delta C$ vary over multiple orders in magnitude. Calculations were performed in CAMB \citep{Howlett2012} using default values from the April 2014 release save for the alteration in the Bose DM background via Eqns. \ref{forceamp} and \ref{pertC}.}
\label{BAOimpact}
\end{center}
\end{figure}

Past and present probes of structure are capable of determining the early power spectra and concentration of gravitating species with percent level precision \citep{Spergel2007,Planck2018Cosmo}. 
Bose dynamics with minor density fluctuations have nearly
indistinguishable dynamics from CDM near the
radiation-matter transition. The linear evolution of large scale
structure formation from a Harrison-Zeldovich primordial power
spectrum produces only weak deviations from CDM for our inflated
condition on the initial correlation, Figs.~\ref{CMBpowspec},
\ref{BAOimpact}.  Linear Bose structures are found to be consistent with the CDM structures to the one percent level for initial correlations in the range $|\delta C| \in [0.0,0.01]$. Structure in the non-linear regime require a more study, which will be applied in the next entries of this series.

\section{Discussion}
\label{Discussion}

The model derived in Section~\ref{New} displays a number of interesting properties. The extremal nature of an exact condensate reveals the nature of inter-axion correlation and exchange in a most natural way, producing new physics on all considered scales. The unique physics appears both in terms of high-order dynamical moments and as novel gravity-inspired potentials, all stemming from the  correlation function $\tilde{g}$. Solutions to the conformal Lagrange multipliers in turn depend on global invariants of the distribution, perhaps due to the unbounded nature of the extremal correlations and exchange symmetry. More realistic depictions of primordial axions are likely to ruin this exceptional case by introduction of new length scales and state transitions, though the current form may be considered to represent the best-case scenario for seeing physics beyond the dissociated limit.

The evolution of linear structures strictly reveals no departure of Bose DM from CDM infall. Deviations enter only at the second order in perturbation and reveal only mild divergences in dynamics and early structure when those liberties are taken; only marginal impacts to current isolated small-scale structures should be expected. This lack of scale dependence is intriguing, and again held to be related to the unbounded nature of the correlation and the non-local exchange symmetry within the region of interest. The marginal result of new structure is not surprising as the non-trivial portion of the correlation function is highly non-linear.

The final unresolved problem for the dynamics of condensed QCD axions is specifying the value of initial correlation. A complete model tracking of quantum correlations from pre-inflation to the
current era is needed to make definitive statements about the presence of axion physics beyond the mean field. Such models, even for a subset of early cosmological history, are currently lacking in the literature.

\section{Summary}
\label{Summary}

The well-motivated axion DM candidate has properties that allow it to simultaneously form a Bose-Einstein condensate and a level of inter-particle correlation unavailable to most bosons studied in material science. This paper sought to derive a robust and computationally effective model of axion infall from first principles, and has done so in the condensed limit. The unique nature of such a maximally condensed axion fluid displays distinct departures from the standard collision-less fluid picture in the relevant case of self-gravitation. These cross-correlation impacts do not appear to significantly influence the outcome of linear structure formation, but may impact structures entering the non-linear regime. The next entries to this series introduce numerical simulation techniques to Bose gravitational collapse and explore the nuances of imperfect and dynamical axion condensation.

\section{Acknowledgments}
\label{Acknowledgments} 

We would like to thank Jens Niemeyer, Katy Clough, David Marsh, Bodo Schwabe, Jan Veltmaat, and Xiaolong Du for their productive discussions in the refinement of this paper. We also gratefully acknowledge the support of the U.S. Department of Energy office of High Energy Physics and the National Science Foundation. TQ was supported in part by the NSF grant AST-1514868. EL and LR were supported in part by the DOE grant DE-SC0011665.

\bibliographystyle{apsrev4-1}
\bibliography{Bibliography.bib}

\appendix
\newpage

\section{The Many-Body Schr\"odinger Equation and One Special Class of Systems}
\label{MBS}

Let us consider the problem of axion infall as a solution of the many-body Schr\"odinger equation under inter-particle Newtonian gravitation.
\begin{equation}
i \hbar \partial_t \Psi \left( \vec{x}_1,...,\vec{x}_N;t \right) = \hat{H} \Psi \left( \vec{x}_1,...,\vec{x}_N;t \right)
\end{equation}
with the Hamiltonian of the form
\begin{align}
\hat{H} &= \hat{T} + \hat{V}_{int} \nonumber \\
&= - \sum_i^N \frac{\hbar^2 \nabla^2}{2 a^2 m_a} + \sum_{i < j}^N \phi_{ij} 
\end{align}
where the external potential from the main text has been ignored. The
domain of the Hamiltonian operator is contained in the Hilbert space
$L^2(\Re^{3N})$, the time operator is spanned by the set of smooth
complex functions, and together $L^2(\Re^{3N})\otimes C_0^1(\Re)$
spans the domain of the Schr\"odinger operator. The product space is
isomorphic to the $(N+1)$-fold tensor product
\begin{equation}
L^2(\Re^{3}) \otimes ... \otimes L^2(\Re^{3}) \otimes C_0^1(\Re)
\end{equation}
as well as the N-body Fock domain 
\begin{equation}
\left(L^2(\Re^{3}) \otimes C_0^1(\Re) \right) \otimes ... \otimes \left(L^2(\Re^{3}) \otimes C_0^1(\Re)\right)/T
\end{equation}
where $T$ is the equal-time equivalence relation among all elements of the product. Being contained in the Fock domain implies any solution of the many-body Schr\"odinger equation may be written as combinations of single-body products to arbitrary accuracy
\begin{equation}
\Psi = \prod_{\substack{i \in \{1,...,N\} \\ \alpha \in A}} \psi_{\alpha} \left( \vec{x}_i \right) + ...
\end{equation}
Finding these combinations can be difficult. With such a strongly correlated force as gravity, these linear combinations are often of infinite length and slowly converging, making computation difficult when faced with finite resources. As a goal of the axion structure formation research project is to succinctly describe the dynamics of the axion system, an additional tool is needed.

As the axions are bosons, the commutation relations of the quantum field requires the system to only occupy fully symmetric states under particle exchange 
\begin{align}
\Psi \left( \vec{x}_1,...,\vec{x}_i,...,\vec{x}_j,...,\vec{x}_N;t \right) &= \Psi \left( \vec{x}_1,...,\vec{x}_j,...,\vec{x}_i,...,\vec{x}_N;t \right) \nonumber \\
&\forall i,j
\end{align}
which further refines the solution domain to
\begin{equation}
\left(L^2(\Re^{3}) \otimes C_0^1(\Re) \right) \otimes ... \otimes \left(L^2(\Re^{3}) \otimes C_0^1(\Re)\right)/(T \circ S_N)
\end{equation}
where $S_N$ is the equivalence among the symmetric group of axion permutations.

To find a more useful and concise form of the many-body solutions, let us rewrite the Hamiltonian in a form that showcases the unique shape of the potential. The kinetic term $\hat{T}$ describes the energy due to motion of axions relative to a prescribed frame of reference, but motion of the collection of particles may just as well be parameterized by the motion of the system's center of mass and the relative particle motions. An appropriate gradient re-parameterization makes this clear 
\begin{align}
\vec{\nabla}_i &= \frac{1}{N} \sum_j^N \left( \vec{\nabla}_j + \left(\vec{\nabla}_i - \vec{\nabla}_j\right)\right) \nonumber \\
&= \frac{1}{N} \vec{\nabla}_{com} + \frac{1}{N}\sum_{i \ne j} \vec{\nabla}_{\Delta_{ij}}
\end{align}
where $\Delta_{ij} = \vec{x}_i - \vec{x}_j$, $\vec{\nabla}_{\Delta_{ij}} = \vec{\nabla}_i - \vec{\nabla}_j$, and $\vec{\nabla}_{com} = \sum_j^N \vec{\nabla}_j $, making
\begin{align}\label{eq:sqprm}
\nabla^2_i &=   \frac{1}{N^2} \nabla^2_{com} + \frac{1}{N^2} \sum_{i \ne j} \nabla^2_{\Delta_{ij}} \\ \nonumber
& + \frac{2}{N^2} \sum_{i \ne j} \vec{\nabla}_{com} \cdot \vec{\nabla}_{\Delta_{ij}} + \frac{1}{N^2} \sum_{k \ne j} \vec{\nabla}_{\Delta_{ij}} \cdot \vec{\nabla}_{\Delta_{ik}} 
\end{align}
The alternate form of $\hat{T}$ can then be written as
\begin{equation}
\hat{T} = -\frac{\hbar^2}{2 a^2 M_{tot}} \nabla^2_{com} - \sum_{i < j} \frac{\hbar^2}{2 a^2 \mu} \nabla^2_{\Delta_{ij}}
\end{equation}
where $\mu = m_a/2N$. The second to last term from Eqn.~\ref{eq:sqprm} is found to vanish from $\hat{T}$ as each $\vec{\nabla}_{com} \cdot \vec{\nabla}_{\Delta_{ij}}$ may be matched with a $\vec{\nabla}_{com} \cdot \vec{\nabla}_{\Delta_{ji}} = - \vec{\nabla}_{com} \cdot \vec{\nabla}_{\Delta_{ij}}$, and the sum of the last term in Eqn.~\ref{eq:sqprm} is found to contribute half of the inter-axion energy. The many-body Hamiltonian can then be rewritten 
\begin{align}
\hat{H} &= -\frac{\hbar^2}{2 a^2 M_{tot}} \nabla^2_{com} - \sum_{i < j} \frac{\hbar^2}{2 a^2 \mu} \nabla^2_{\Delta_{ij}} - \sum_{i < j}^N \phi_{ij}
\end{align}
and can be separated into a center-of-mass Hamiltonian and an interacting Hamiltonian 
\begin{align}
\hat{H}_{com} &= -\frac{1}{2 M_{tot}} \nabla^2_{com} \\
\hat{H}_{int} &= - \sum_{i < j} \frac{1}{2 \mu} \nabla^2_{\Delta_{ij}} + \sum_{i < j}^N \phi_{ij}
\end{align}

Note that $\hat{H}_{int}$ appears to be separable over the
$\vec{\Delta}_{ij}$, though the $\Delta$s are non-independent over the
many-body solution space. To circumvent this, let use consider a
slightly different domain: a system of $N(N-1)$ `particles' with
Hamiltonian $H_{int}$ defined over the function space
$L^2(\Re^{3N(N-1)})\otimes C_0^1(\Re)$ with an analogous Bose constraint 
\begin{equation}
L^2(\Re^{3N(N-1)})\otimes C_0^1(\Re)/S_{N(N-1)}
\end{equation}
and all the implications thereof. In this case the Hamiltonian is separable and the stationary states are spanned by single permanents of single-`particle' stationary states
\begin{equation}
\Phi' \left( \Delta_{ij} \right) \in sp \left( \left\{ perm \left( \prod_{ij } \phi_{\alpha_{ij}} \left(\vec{x}_i - \vec{x}_j,t \right) \right) \right\}_{\alpha \in A} \right)
\end{equation}
and the pure states as 
\begin{equation}
\Psi' \left( \Delta_{ij}, t \right) \in sp \left( \left\{ perm \left( \prod_{ij } \psi_{\alpha_{ij}} \left(\vec{x}_i - \vec{x}_j,t \right) \right) \right\}_{\alpha \in A} \right) 
\end{equation}
where $sp()$ gives the linear span of a collection of functions, $perm()$ is the permanent operation, $A$ is the set of many-body solution indexes, $\hat{\psi}_{a}$ is a solution to the single-body equation
\begin{equation}
i \hbar \partial_t \psi_{a} = - \frac{\hbar^2}{2 a^2 \mu} \nabla^2_{\left( \vec{x}_i - \vec{x}_j \right)} \psi_{\alpha} + \phi\left( \vec{x}_i - \vec{x}_j \right) \psi_{a}
\end{equation}
Solutions of this $N(N-1)$ space may be used in the original use-case of $N$ axions by applying an appropriate projection
\begin{equation}
p: \text{ } \Delta_{ij} = \vec{x}_i - \vec{x}_j  \text{ }  \forall i,j
\end{equation}
on the Hamiltonian and function space pair.
\begin{align}
&\left( \hat{H}\left( \Re^{3 N(N-1) +1} \right), \mathscr{C}\left( \Re^{3 N(N-1) +1} \right) \right) \nonumber \\
&\to \left( \hat{H}\left( \Re^{3 N +1} \right), \mathscr{C}\left( \Re^{3 N +1} \right) \right)
\end{align}
It is apparent that this projection both respects the modulated many-body solution space and the N-body Schr\"odinger equation. It can also be shown that the span of solutions to the $N$-body problem can be represented by projected solutions of the $N(N-1)$-body problem, implying $p\left(L^2(\Re^{3N(N-1)})\otimes C_0^1(\Re)/S_{N(N-1)} \right)$ is dense in $L^2(\Re^{3N})\otimes C_0^1(\Re)/S_{N}$.

\section{The Many-Body Wigner Function and its Super-de Broglie Limit}
\label{Wigner}

This is a short review of the many-body Wigner transform, some of its qualities, and its equation of motion in the super-de Broglie limit.

\subsection{Wigner Transform}

Phase space DFs are often far better suited for tracking system observables than wave-functions, due to the hyperbolicity of their governing dynamics. The many-body Bose system can be reduced into a phase-space pseudo-DF via the many-body Wigner function
\begin{align}
&f^{(N)}(\vec{x}_1,\vec{p}_1,...,\vec{x}_N,\vec{p}_N,t) \nonumber \\
&= \int d^3x'_1 \cdot \cdot \cdot d^3x'_N e^{i \sum_j^N \vec{p}_j \cdot \vec{x}'_j/\hbar} \times \nonumber \\
& \Psi^{\dagger}(\vec{x}_1+\vec{x}'_1/2,...,\vec{x}_N+\vec{x}'_N/2,t) \times \nonumber \\
&\Psi(\vec{x}_1-\vec{x}'_1/2,...,\vec{x}_N-\vec{x}'_N/2,t)
\end{align}
For brevity, the arguments of $\Psi^{\dagger}$ and $\Psi$ are understood to be $\{\vec{x}_i + \vec{x}_i'/2\}$ and $\{\vec{x}_i - \vec{x}_i'/2\}$ respectively. While not a true DF, in that $f^{(N)}$ can take on negative values,
it does share a number of properties with true DFs, such as being real-valued
\begin{align}
f^{(N)}(\vec{x},\vec{p})^* &= \int d^3x' e^{-i \sum_j^N \vec{p}_j \cdot \vec{x}'_j/\hbar} \left(\Psi^{\dagger}  \Psi \right)^*\nonumber \\
&= \int d^3x' e^{-i \sum_j^N \vec{p}_j \cdot \vec{x}'_j/\hbar} \Psi  \Psi^{\dagger} \nonumber \\
&= \int d^3x' e^{i \sum_j^N \vec{p}_j \cdot \vec{x}'_j/\hbar} \Psi^{\dagger}  \Psi \text{, }(\vec{x}' \to -\vec{x}') \nonumber \\
&= f^{(N)}(\vec{x},\vec{p}),
\end{align}
and providing properly-normalized probability densities over projections to 3D single-body spaces
\begin{align}
\rho(\vec{x}) &= \int d^3 p_1 d^{6(N-1)} w f^{(N)}, \\
\rho(\vec{p}) &= \int d^3 x_1 d^{6(N-1)} w f^{(N)}.
\end{align}
The Wigner pseudo-DF can also be expected to provide a reliable DF in certain circumstances \citep{Zurek2003,Zachos2005,Bondar2013}, namely on scales much larger than the de Broglie wavelength. The differences between a Wigner function and a true DF stem from the uncertainty principles present in the underlying wave-function/quantum-state representation. Complementary parameters such as position and momentum cannot be fully articulated simultaneously when quantum limited, but we see soon that the two become independent at the effective structure formation scales, far above the de Broglie length. We choose the DF form as it fully captures the dynamics of diffusive quantum systems and is most convenient for N-Body simulation, a numerical technique approached later in this series.

\subsection{Super de-Broglie Equation of Motion}

The equation of motion for the distribution is found by straightforward differentiation
\begin{align}
&\partial_t f^{(N)}(\vec{x}_1,\vec{p}_1,...,\vec{x}_N,\vec{p}_N,t) \nonumber \\
&= \int d^3x'_1 \cdot \cdot \cdot d^3x'_N e^{i \sum_j^N \vec{p}_j \cdot \vec{x}'_j/\hbar} \left(\partial_t \Psi^{\dagger}  \Psi + \Psi^{\dagger} \partial_t \Psi \right)
\end{align} 
Judicious substitution of the Schr\"odinger equation transforms the expression into an equation of motion
\begin{align}
&\partial_t f^{(N)}(\vec{x}_1,\vec{p}_1,...,\vec{x}_N,\vec{p}_N,t) \nonumber \\
&= \frac{1}{i \hbar} \int d^3x'_1 \cdot \cdot \cdot d^3x'_N e^{i \sum_j^N \vec{p}_j \cdot \vec{x}'_j/\hbar} \left(- \hat{H} \Psi^{\dagger}  \Psi + \Psi^{\dagger} \hat{H} \Psi \right)
\end{align}
where we use the Hamiltonian of Eqn.\ref{MBHam}.

The kinetic contribution can be simplified through successive application of the product rule of differentiation and consolidation of terms, and as expected provides transport terms
\begin{align}
&K = \frac{1}{i \hbar}\frac{\hbar^2}{2 a^2 m} \sum_i^N \int d^3x'_1 \cdot \cdot \cdot d^3x'_N e^{i \sum_j^N \vec{p}_j \cdot \vec{x}'_j/\hbar} \times \nonumber \\
&\left(\nabla^2_{(\vec{x}_i+\vec{x}'_i/2)}  \Psi^{\dagger}  \Psi - \Psi^{\dagger} \nabla^2_{(\vec{x}_i-\vec{x}'_i/2)} \Psi \right) \nonumber \\
&= - \sum_i^N \frac{\vec{p}_i}{a^2 m} \vec{\nabla}_i f^{(N)} + O(\hbar)
\end{align}
where de Broglie wavelength contributions  and higher order are inconsequential. The quantum diffusion scales of the QCD axion are many orders of magnitude smaller than the desired scale of simulation, causing them to average out to zero over cosmological coarse graining.

The external potentials give forcing contributions, calculated through Taylor expansion of the potentials about $\vec{x}_i = 0$:
\begin{align}
F_{ext} &= \frac{1}{i \hbar} \sum_i^N \int d^3x'_1 \cdot \cdot \cdot d^3x'_N e^{i \sum_j^N \vec{p}_j \cdot \vec{x}'_j/\hbar} \times \nonumber \\
&\left(\Phi' (\vec{x}_i+\vec{x}'_i/2)  \Psi^{\dagger}  \Psi - \Psi^{\dagger} \Phi' (\vec{x}_i-\vec{x}'_i/2) \Psi \right) \nonumber \\
&= \sum_i^N \vec{\nabla}_i \Phi' \cdot \vec{\nabla}_{p_i} f^{(N)}  + O(\hbar^2)
\end{align}
Lastly, the two-body interaction potentials also contribute a forcing term
\begin{align*}
&F_{int} = \frac{1}{2 i \hbar} \sum_{i \ne j}^N \int d^3x'_1 \cdot \cdot \cdot d^3x'_N e^{i \sum_j^N \vec{p}_j \cdot \vec{x}'_j/\hbar} \times \nonumber \\
&\Big( \phi_{ij} (\vec{x}_i+\vec{x}'_i/2,\vec{x}_j+\vec{x}'_j/2)  \Psi^{\dagger}  \Psi \nonumber \\
&- \Psi^{\dagger} \phi_{ij} (\vec{x}_i-\vec{x}'_i/2,\vec{x}_j-\vec{x}'_j/2) \Psi \Big) \nonumber \\
&= \sum_{i \ne j}^N \vec{\nabla}_i \phi_{ij} \cdot \vec{\nabla}_{p_i} f^{(N)}  + O(\hbar^2)
\end{align*}
recalling that $\phi_{ij}$ is the two-body gravitational potential. The result is a Liouville-like equation of motion that appears very similar to the classical form for a collision-less fluid
\begin{align}
&\partial_t f^{(N)} = K + F_{ext} + F_{int} \\
&= - \sum_i^N \frac{\vec{p}_i}{a^2 m} \vec{\nabla}_i f^{(N)} + \sum_i^N \vec{\nabla}_i \Phi' \cdot \vec{\nabla}_{p_i} f^{(N)}  \nonumber \\
&+ \frac{1}{2}\sum_{i \ne j}^N \vec{\nabla}_i \phi_{ij} \cdot \vec{\nabla}_{p_i} f^{(N)} + O(\hbar) \label{DFEOM}
\end{align}
We concern ourselves with the Liouville-like equation composed of only lowest order terms. Note that so far we have not enforced specific dynamics or even exchange conditions on the state. Such a Boltzmann equation may also by applied to analogous classes of fermionic or mixed systems.

\section{Projection of the Many-Body Distribution Function}
\label{DFInt}

Projection of the many-body DF down to the total phase space density is accomplished through integration of phase spaces $2,3,...,N$. For general quantum and classical systems, correlations in the (psuedo-)DF $f^{(N)}$ may be present between any subset of particles. Fortunately, the topic system provides us with insight to the correlations in the form of a complete basis of solutions to the governing Schr\"odinger equation.

\subsection{Reduction to Single DF}
\label{}

We follow the well-used approach of density projection by integrating out the supurfluous degrees of freedom
\begin{equation}
f^{(M)} = \int d^6 w_{M+1} \cdot \cdot \cdot d^6 w_N f^{(N)}
\end{equation}
where $w_i = ( \vec{x}_i, \vec{p}_i )$ is the i-th six dimensional phase space. Projection of the governing super de Broglie Liouville equation
\begin{align}
&\partial_t f^{(N)} + \sum_i^N \frac{\vec{p}_i}{a^2 m_a} \vec{\nabla}_i f^{(N)} - \sum_i^N \vec{\nabla}_i \Phi' \cdot \vec{\nabla}_{p_i} f^{(N)}  \nonumber \\
&- \frac{1}{2}\sum_{i \ne j}^N \vec{\nabla}_i \phi_{ij} \cdot \vec{\nabla}_{p_i} f^{(N)} = O(\hbar) \label{DFEOM}
\end{align}
is expected to follow analogously.

Performing the single-body projection on the lowest order of the  Liouville equation
\begin{eqnarray}
&0 = \int d^6 w_2 \cdot \cdot \cdot d^6 w_N \times \nonumber \\
& \left[ \partial_t f^{(N)} + \sum_i^N \frac{\vec{p}_i}{m} \vec{\nabla}_i f^{(N)} - \sum_i^N \vec{\nabla}_i \Phi' \cdot \vec{\nabla}_{p_i} f_N \right. \nonumber \\
& \left. - \frac{1}{2} \sum_{i \ne j}^N \vec{\nabla}_i \phi_{ij} \cdot \vec{\nabla}_{p_i} f^{(N)}   \right]
\end{eqnarray}
reduces the explicit time component to
\begin{equation}
T^{(1)} \equiv \int d^6 w_2 \cdot \cdot \cdot d^6 w_N \partial_t f^{(N)} = \partial_t f^{(1)},
\end{equation}
The kinetic transport term reduces as expected, due to the divergence theorem and axion conservation
\begin{equation}
K^{(1)} \equiv \int d^6 w_2 \cdot \cdot \cdot d^6 w_N \sum_i^N \frac{\vec{p}_i}{a^2 m} \vec{\nabla}_i f^{(N)} = \frac{\vec{p}}{a^2 m_a} \vec{\nabla} f^{(1)} ,
\end{equation}
The external potential forcing term reduces similarly
\begin{equation}
F_{ext}^{(1)} \equiv - \int d^6 w_2 \cdot \cdot \cdot d^6 w_N \sum_i^N \vec{\nabla}_i \Phi' \cdot \vec{\nabla}_{p_i} f^{(N)} = - \vec{\nabla} \Phi' \cdot \vec{\nabla}_{p} f^{(1)} 
\end{equation}
It is the self-gravity forcing term that is the most troublesome, as the two-body gravitational interactions contain a measure of two-body correlations
\begin{align}
&F_{int}^{(1)} \equiv - \frac{1}{2} \int d^6 w_2 \cdot \cdot \cdot d^6 w_N \sum_{i \ne j}^N \vec{\nabla}_i \phi_{ij} \cdot \vec{\nabla}_{p_i} f^{(N)} \nonumber \\
&= - \int d^6 w_j \sum_{j > 1}^N \vec{\nabla} \phi_{1j} \cdot \vec{\nabla}_{p} f^{(2)} \label{DFFint}
\end{align}

\subsection{Correlation  Hints} 
\label{}

Let us take stock of correlations in the many-body distribution function before attempting further simplification. Recall that the considered many-body DF may be expressed in terms of a single condensed state 
\begin{align}
&f^{(N)}(\vec{x}_1,\vec{p}_1,...,\vec{x}_N,\vec{p}_N,t) \nonumber \\
&= \int d^3x'_1 \cdot \cdot \cdot d^3x'_N e^{i \sum_j^N \vec{p}_j \cdot \vec{x}'_j/\hbar} \nonumber \\
&\prod_{i \ne j} \hat{\psi}^{\dagger}_{a} \left(\vec{x}_i - \vec{x}_j + \vec{x}'_i/2 - \vec{x}'_j/2,t \right) \nonumber \\
&\times \prod_{i \ne j} \hat{\psi}_{a} \left(\vec{x}_i - \vec{x}_j - \vec{x}'_i/2 + \vec{x}'_j/2,t \right)
\end{align}
where the condensed state is of the form derived in Appx~\ref{MBS}. Condensate wave-functions comprised solely of two-axion correlators conspire with the reach of inter-axion gravity to expose that two-body correlations exist between axions and that they are the only correlations relevant to the formation of large scale structure. Finally, the long-range nature of exchange symmetry in this NR co-moving context indicates that the correlation function between axions should be intrinsically scale independent.

\subsection{Correlation Function}

To derive the form of two-body correlations relevant for large scale
structure, we first rewrite the two-body DF, $f^{(2)}$, from Eqn.~\ref{DFFint} in a form that makes the distinction between correlations and total density more obvious
\begin{equation}
f^{(2)}(w_1,w_2,t) = \tilde{g} \times f^{(1)}(w_1,t) f^{(1)}(w_2,t) 
\end{equation}
where the correlation function $\tilde{g}$ is seen to be symmetric over both phase spaces. 

This correlation form has some convenient properties, including
converging with standard definitions of correlation in the condensed
matter literature in the limit of $N \to \infty$. In addition to the
standard normalization of the single-body DF,
\begin{equation}
\int d^6 w_1 f^{(1)} = 1,
\end{equation} 
there is also a normalization condition on the correlation function,
\begin{equation}
\int d^6 w_1 \tilde{g} f^{(1)} = 1. \label{corrconstr}
\end{equation}

This correlation function is not expected to be of great computational utility as an explicitly wave-function expression, though it can be shown that the totally condensed axion system inhabits an extremal correlation state in the sense that the functional
\begin{equation}
I = \int d^6w_1 d^6w_2 \tilde{g} 
\end{equation}
has vanishing physical variation about the condensed state ($\delta I = 0$).  A modified functional may be constructed 
\begin{equation}
I' = I + \lambda_1\left(\int d^6 w_1 f^{(1)} - 1\right) + \lambda_2\left(\int d^6 w_1 \tilde{g} f^{(1)} -1\right) \label{corr_extr}
\end{equation}
using correlation-normalization Lagrange multipliers, constraining the correlator to physicality. Beyond this point we continue the derivation using physical conjectures due to the intractability of evaluating the full many-body expressions.

We conjecture that $\tilde{g}$'s dependence on $f^{(1)}$ goes as
\begin{equation}
f_+ = \frac{1}{2}\left(f^{(1)}(w_1,t) + f^{(1)}(w_2,t) \right).
\end{equation}
The symmetric nature of $\tilde{g}(f_+)$ implies that the extremization of Eqn. \ref{corr_extr} is equivalent to extremization of
\begin{equation}
I'' = \int d^6w_1 \tilde{g} + \lambda_1\left(\int d^6 w_1 f^{(1)} - 1\right) + \lambda_2\left(\int d^6 w_1 \tilde{g} f^{(1)} -1\right) 
\end{equation}
This functional can be simply extremized over $f^{(1)}$ to find the condensed $\tilde{g}$. Variational principles show that the extremals of $I''$ satisfy
\begin{equation}
\frac{\partial \tilde{g}}{\partial f^{(1)}} + \lambda_1 + \lambda_2
\left(\tilde{g} + f^{(1)} \frac{\partial \tilde{g}}{\partial f^{(1)}}
\right) = 0.
\end{equation}
The symmetry of $\tilde{g}$ over phase spaces also allows for symmetrization of the extremal characteristic equation, resulting in the equivalent extremization condition
\begin{equation}
\frac{\partial \tilde{g}}{\partial f_+} + \lambda_1 + \lambda_2
\left(\tilde{g} + f_+ \frac{\partial \tilde{g}}{\partial f_+} \right)
= 0.
\end{equation}
Solutions to the condition are found to be
\begin{equation}
\tilde{g} = \frac{C-\lambda_1 f_+}{1+ \lambda_2 f_+}
\end{equation}
where $C$ is constrained by the correlation present in the initial configuration. For instance, a separable Bose condensate gives $C=1$, with Lagrange multipliers also conforming to the uncorrelated dynamics of standard MFTs. 

\begin{figure}[]
\begin{center}
\includegraphics[width=9cm]{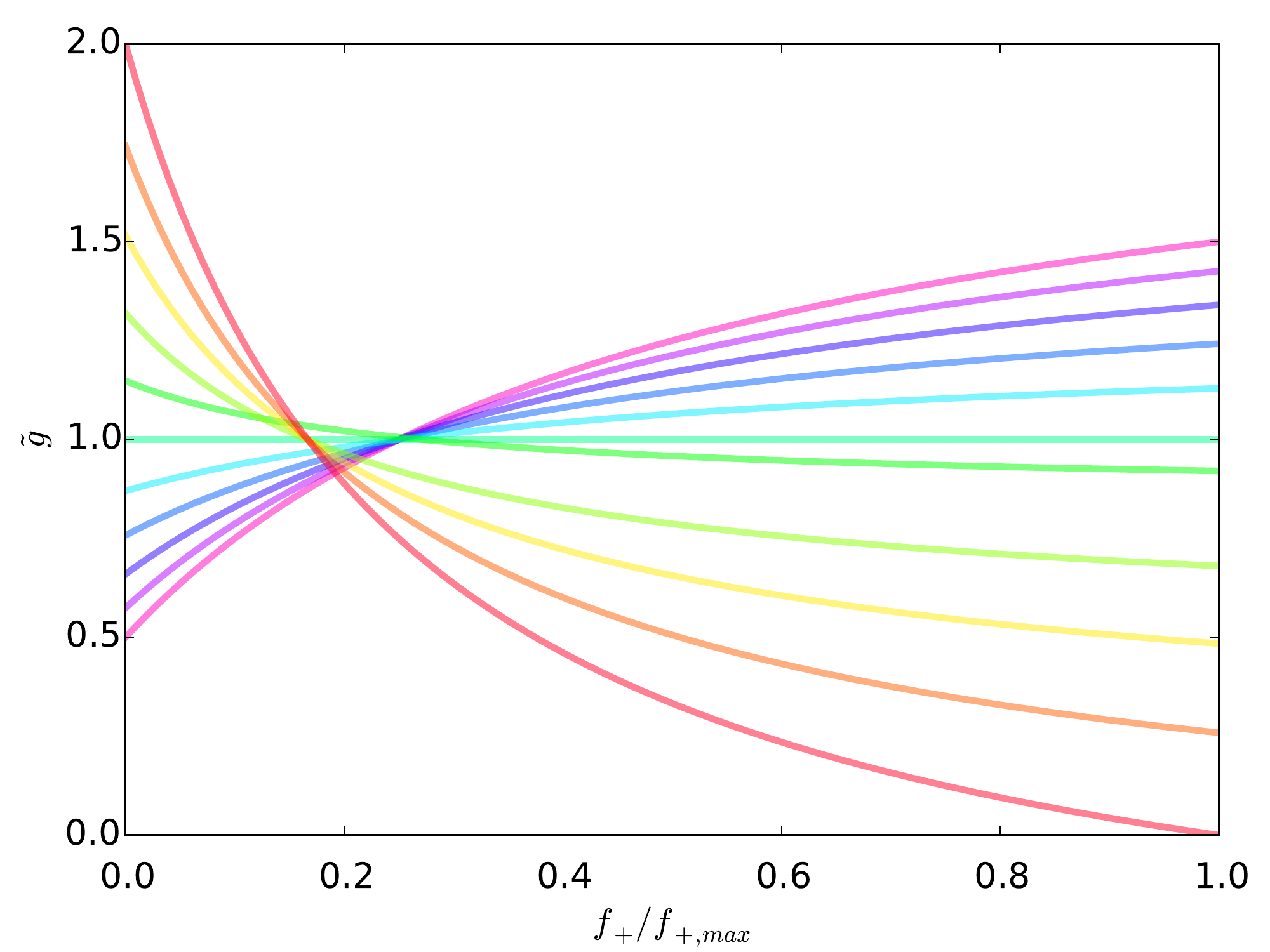}
\caption{ Correlation functions for several types of systems, both bosonic and fermionic. The lines that lie below unity at $f_+ = f_{+,max}$ (red to green) are from non-interacting Fermion distributions, organized in a spin-symmetric configuration, parameterized by temperature decreasing with initial correlation like $T = T_f/(C-1)^2$, where $T_f$ is the Fermi temperature. The lines at and above unity (teal to violet) are from condensed scalar Boson distributions shaped into a spherical Gaussian in phase space, and also become colder with more extreme initial correlation via the dispersion length relation $\tilde{\sigma}_p/\tilde{\sigma}_x \propto C$.}
\label{gplot}
\end{center}
\end{figure}

The extremal values of $\lambda_1$ and $\lambda_2$ may be found through the constraint equations. The correlation constraint, Eqn. \ref{corrconstr}, for the extremal solutions becomes
\begin{equation}
1 = \int d^6 w_1f^{(1)}(w_1,t) \frac{C-\lambda_1 f_+}{1+ \lambda_2 f_+}.
\end{equation}
See Fig.~\ref{gplot} for an example of the behavior of this correlator.

The forcing term of Eqn.\ref{DFFint} may now be written 
\begin{align}
F_{int} &= - \int d^6w_j \sum_{j > 1}^N \vec{\nabla}_1 \phi_{1j} \cdot \vec{\nabla}_{p_1} f^{(2)} = \nonumber \\
&= - \int d^6w_j \sum_{j > 1}^N \vec{\nabla}_1 \phi_{1j} \cdot \vec{\nabla}_{p_1} f^{(1)}(w_1,t) f^{(1)}(w_j,t) \nonumber \\
&\times \frac{C-\lambda_1 f_+}{1+\lambda_2 f_+} \label{fintfinal}
\end{align}
producing an new collision-less Boltzmann-like equation 
\begin{align}
&0=\partial_t f^{(1)} + \frac{\vec{p}}{m} \cdot \vec{\nabla} f^{(1)} - \vec{\nabla} \Phi' \cdot \vec{\nabla}_p f^{(1)} \nonumber \\
& - \frac{N}{N-1}\int d^6w_j \vec{\nabla} \phi_{1j} \cdot \vec{\nabla}_{p} f^{(1)}(w_1,t) f^{(1)}(w_j,t) \nonumber \\
&\times \frac{C-\lambda_1 f_+}{1+\lambda_2 f_+}
\end{align}
with the last term containing both the Newtonian and the exchange-correlation contributions of inter-axion gravitation.

\section{Expansions and Limits of the Condensed Boltzmann Equation}
\label{EXP}

The full form of the Boltzmann-like equation for condensed Bose systems found in
Appx.~\ref{DFInt} is quite compact, though difficult to use
analytically due to the presence of non-polynomial
non-linearities. Semi-analytic techniques, such as the early universe
structure formation calculations of Section \ref{AS}, would benefit
from an expansion in orders.  We provide the first few orders here.

Expanding $F_{int}$ from Eqn.~\ref{fintfinal} in orders of $\lambda_2 f_+$ and organizing in terms of powers of $f^{(1)}$ produces
\begin{align*}
F_{int} &= - \int d^6w_j \sum_{j > 1}^N \vec{\nabla}_1 \phi_{1j} \cdot \vec{\nabla}_{p_1} f^{(1)}(w_1,t) f^{(1)}(w_j,t) \nonumber \\
& \times \left(C-\lambda_1 f_+\right)\sum_{n=0}^{\infty} \left(-\lambda_2 f_+\right)^n \nonumber \\
&= -(N-1) C \vec{\nabla} \bar{\Phi} \cdot \vec{\nabla}_p f^{(1)} \nonumber \\
&+(N-1)\frac{\lambda_1+\lambda_2}{2} \Biggl( \vec{\nabla} \bar{\Phi} \cdot \vec{\nabla}_p \left(f^{(1)}\right)^2 \nonumber \\
&+ \vec{\nabla} \int d^6w_2 \phi_{12} \left(f^{(1)}\right)^2 \cdot \vec{\nabla}_p f^{(1)}  \Biggr) \nonumber \\
&-(N-1)\frac{\lambda_2(\lambda_1+\lambda_2)}{4} \Biggl(  \vec{\nabla} \bar{\Phi} \cdot \vec{\nabla}_p \left(f^{(1)}\right)^3 \nonumber \\
&+ 2\vec{\nabla} \int d^6w_2 \phi_{12} \left(f^{(1)}\right)^2 \cdot \vec{\nabla}_p \left(f^{(1)}\right)^2 \nonumber \\
&+ \vec{\nabla} \int d^6w_2 \phi_{12} \left(f^{(1)}\right)^3 \cdot \vec{\nabla}_p f^{(1)}  \Biggr) \nonumber \\
&+(N-1)\frac{\lambda_2^2(\lambda_1+\lambda_2)}{8} \Biggl( \vec{\nabla} \bar{\Phi} \cdot \vec{\nabla}_p \left(f^{(1)}\right)^4 \nonumber \\
&+ 3 \vec{\nabla} \int d^6w_2 \phi_{12} \left(f^{(1)}\right)^2 \cdot \vec{\nabla}_p \left(f^{(1)}\right)^3 \nonumber \\
&+ 3 \vec{\nabla} \int d^6w_2 \phi_{12} \left(f^{(1)}\right)^3 \cdot \vec{\nabla}_p \left(f^{(1)}\right)^2 \nonumber \\
&+ \vec{\nabla} \int d^6w_2 \phi_{12} \left(f^{(1)}\right)^4 \cdot \vec{\nabla}_p f^{(1)}  \Biggr) \nonumber \\
&+ O\left(\left(f^{(1)}\right)^6\right)
\end{align*}
The governing Boltzmann-like equation then expands to
\begin{align}
&0=\partial_t f^{(1)} + \frac{\vec{p}}{m} \cdot \vec{\nabla} f^{(1)} - \vec{\nabla} \Phi' \cdot \vec{\nabla}_p f^{(1)} -\vec{\nabla} \bar{\Phi} \cdot \vec{\nabla}_p f^{(1)} \nonumber \\
&+\frac{\lambda_1+\lambda_2}{2} \Biggl( \vec{\nabla} \bar{\Phi} \cdot \vec{\nabla}_p \left(f^{(1)}\right)^2 \nonumber \\
&+ \vec{\nabla} \int d^6w_2 \phi_{12} \left(f^{(1)}\right)^2 \cdot \vec{\nabla}_p f^{(1)}  \Biggr) \nonumber \\
&-\frac{\lambda_2(\lambda_1+\lambda_2)}{4} \Biggl(  \vec{\nabla} \bar{\Phi} \cdot \vec{\nabla}_p \left(f^{(1)}\right)^3 \nonumber \\
&+ 2\vec{\nabla} \int d^6w_2 \phi_{12} \left(f^{(1)}\right)^2 \cdot \vec{\nabla}_p \left(f^{(1)}\right)^2 \nonumber \\
&+ \vec{\nabla} \int d^6w_2 \phi_{12} \left(f^{(1)}\right)^3 \cdot \vec{\nabla}_p f^{(1)}  \Biggr) \nonumber \\
&+\frac{\lambda_2^2(\lambda_1+\lambda_2)}{8} \Biggl( \vec{\nabla} \bar{\Phi} \cdot \vec{\nabla}_p \left(f^{(1)}\right)^4 \nonumber \\
&+ 3 \vec{\nabla} \int d^6w_2 \phi_{12} \left(f^{(1)}\right)^2 \cdot \vec{\nabla}_p \left(f^{(1)}\right)^3 \nonumber \\
&+ 3 \vec{\nabla} \int d^6w_2 \phi_{12} \left(f^{(1)}\right)^3 \cdot \vec{\nabla}_p \left(f^{(1)}\right)^2 \nonumber \\
&+ \vec{\nabla} \int d^6w_2 \phi_{12} \left(f^{(1)}\right)^4 \cdot \vec{\nabla}_p f^{(1)}  \Biggr) \nonumber \\
&+ O\left(\left(f^{(1)}\right)^6\right)
\end{align}
where we have taken $N-1$ to be indistinguishable from $N$ in cosmological contexts. Energy, momentum, and angular momentum remain conserved at each level of expansion and in the full correlation form.

Solutions to the Lagrange multipliers may also be approximated in some cases, using a geometric expansion of the characteristic constraint equation
\begin{equation}
1 = \int d^6 w_1f^{(1)}(w_1,t) \frac{C-\lambda_1 f_+}{1+ \lambda_2 f_+} \nonumber
\end{equation}
assuming values of $| \lambda_2 f_+|$ not exceeding unity over the region of integration. The functional condition is then translated into a sequence of quasi-algebraic conditions
\begin{align}
& \delta_{k0} = \nonumber \\
& \int d^6w_1 f^{(1)} \left(C- \lambda_1 f^{(1)}/2 \right) \sum_{n \ge k}^{\infty} {{n}\choose{k}} \left(\frac{-\lambda_2}{2}\right)^n \left(f^{(1)}\right)^{n-k}  \nonumber \\
& - \int d^6w_1 f^{(1)} \lambda_1f^{(1)}/2 \sum_{n \ge k-1}^{\infty} {{n}\choose{k-1}} \left(\frac{-\lambda_2}{2}\right)^n \left(f^{(1)}\right)^{n-(k-1)} \label{constrexp}
\end{align}
for $k=0,1,2,...$. Only $k=0$ and one other equation are linearly independent. Here $k=0,1$ are chosen to be solved. The infinite order of the polynomials brings another challenge, so instead we settle for a truncated solution as these sums are also convergent under the $|\lambda_2 f|$ ansatz. Order of the self-expectation values of $f^{(1)}$
\begin{equation}
\bar{f^l} \equiv \int d^6w_1 \left(f^{(1)}\right)^{l+1}
\end{equation}
are chosen as the expansion parameter, with the normalization constraint setting $\bar{f^0}=1$. The first few powers are analytically solvable, though in general a numerical scheme is needed to solve to arbitrary order.

\end{document}